\def\ube13{UBe$\rm_{13}$}

\def\bi2212{Bi$\rm_2$Sr$\rm_2$CaCu$\rm_2$O$\rm_8$}
\def\ybi2212{Bi$\rm_2$Sr$\rm_2$YCu$\rm_2$O$\rm_8$}
\def\ycabi2212{Bi$\rm_2$Sr$\rm_2$Ca$\rm_{1-x}$Y$\rm_x$Cu$\rm_2$O$\rm_{8+\delta}$}

\def\y65cabi2212{Bi$\rm_2$Sr$\rm_2$Ca$\rm_{0.35}$Y$\rm_{0.65}$Cu$\rm_2$O$\rm_{8+\delta}$}
\def\Ir{CeIrIn$_5$}
\def\Co{CeCoIn$_5$}

\def\IrRh{CeIr$_{1-x}$Rh$_x$In$_5$}

\documentstyle[aps,prl,twocolumn,epsf]{revtex}
\begin{document} 
\draft

\def\dfrac#1#2{{\displaystyle{#1\over#2}}}
\twocolumn[\hsize\textwidth\columnwidth\hsize\csname @twocolumnfalse\endcsname
\hfill{Submitted to PRB Rapid Commun.}

\title{On the origin of the zero-resistance anomaly in heavy fermion superconducting \Ir: a clue from magnetic field and Rh-doping studies}
\author{A. Bianchi, R. Movshovich, M. Jaime, J. D. Thompson, P. G. Pagliuso, and J. L. Sarrao}
\address{Los Alamos National Laboratory, Los Alamos, New Mexico 87545}

\date{\today}

\maketitle

\begin{abstract} 

We present the results of the specific heat and AC magnetic susceptibility measurements of \IrRh\ for x from 0 to 0.5. As x is increased from 0 both quantities reflect the competition between two effects. The first is a suppression of superconductivity below the bulk transition temperature of T$_c = 0.4$ K, which is due to the pair breaking effect of Rh impurities. The second is an increase in the volume fraction of the superconducting regions above T$_c$, which we attribute to defect-induced strain. Analysis of the H-T phase diagram for \Ir\ obtained from the bulk probes and resistance measurements points to the filamentary origin of the inhomogeneous superconductivity at T$_\rho \approx 1.2$ K, where the resistance drops to zero. The identical anisotropies in the magnetic field dependence of the specific heat and the resistance anomalies in \Ir\ indicate that the filamentary superconductivity is intrinsic, involving electrons from the part of the Fermi surface responsible for bulk superconductivity.

\end{abstract}


 ]
\narrowtext

Recently discovered ambient-pressure heavy-fermion superconductivity in \Ir~\cite{petrovic:epl_01} is the subject of intense experimental and theoretical investigations. This interest was sparked by an opportunity to study a Ce-based heavy fermion superconductor at ambient pressure - \Ir\ being the second such compound after CeCu$_2$Si$_2$~\cite{steglich79:cecu2si2}, discovered more than two decades ago. A number of U-based heavy fermion superconductors have been shown unambiquously to be unconventional, including UPt$_3$~\cite{fisher:prl_89} and UPd$_2$Al$_3$~\cite{tou:jpsj_95}. \Ir\ and \Co\ (also an ambient pressure heavy fermion superconductor~\cite{petrovic:jpcm_01}) present an opportunity to study in detail the mechanism of the heavy fermion superconductivity in Ce-based compounds. 

\begin{figure}
\epsfxsize=3in
\centerline{\epsfbox{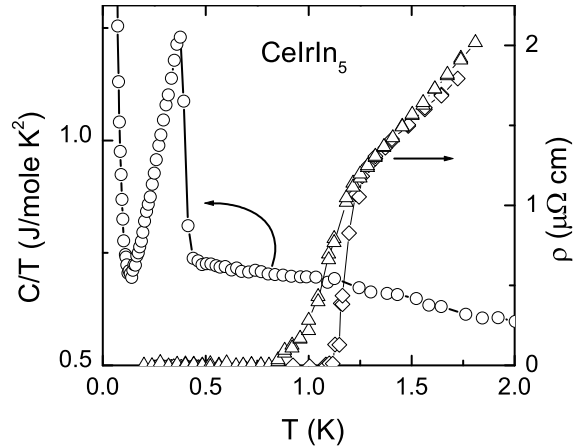}}
\caption{Specific heat C and resistivity $\rho$ \Ir.  ($\circ$) C;  ($\diamond$) $\rho$ of sample \#1; ($\bigtriangleup$) $\rho$ of  sample \#2.} 
\label{R and C}
\end{figure}

Superconductivity in \Ir\ was quickly confirmed by other groups in the US~\cite{stewart:prb_01} and Japan~\cite{haga:unpublished_00,zheng:prl_01,kohori:preprint_00}. The overall properties of \Ir\ samples do not depend on the place of origin, due to the straightforward sample growth procedure and stability of the nominal stoichiometry. The specific heat anomaly (see Fig.~\ref{R and C}, some of these data were also shown in Ref.~\cite{petrovic:epl_01}) indicates bulk superconducting transition at T$_c = 0.4$ K. The resistivity data for two samples are also plotted in Fig.~\ref{R and C}, and show the resistance going to zero at T$_\rho \approx 1$ K. A similar discrepancy was observed in all of the synthesized samples of \Ir, both ours and those made by other researchers. We have already learned a lot about the bulk properties of the superconducting state at low temperature. Both NQR~\cite{zheng:prl_01} and specific heat and thermal conductivity~\cite{movshovich:prl_01} measurements indicate that superconductivity in \Ir\ is unconventional, with lines of nodes in the energy gap. In spite of the accumulated knowledge of the properties of \Ir\ in the superconducting state, the observation of the discrepancy between the superconducting transition temperature T$_c$ and T$_\rho$ remains unresolved.  It has been suggested that the resistance anomaly in \Ir\ may be similar to the formation of the pseudogap in High Temperature Superconductors~\cite{petrovic:epl_01}. Another interpretation suggested the existance of a small amount of second phase with superconducting transition at T$_\rho = 1.2$ K, extrinsic to \Ir. Clearly, it is important to resolve this mystery.

\begin{figure}
\epsfxsize=3in
\centerline{\epsfbox{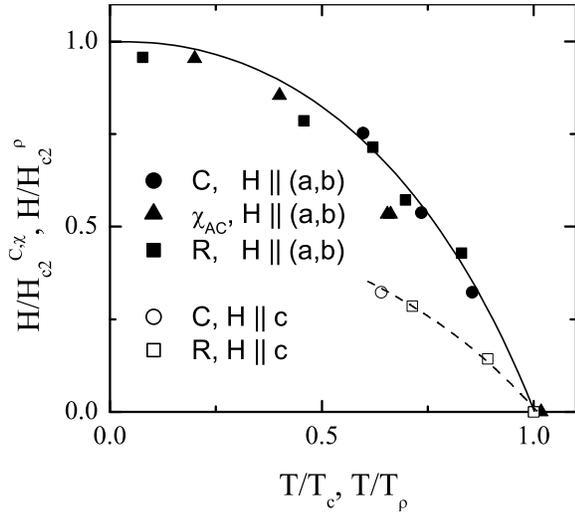}}
\caption{Scaled H-T phase diagram of \Ir. The field H and temperature T of the resistance data are divided by H$_{c2}^\rho = 7$ T ($\mathbf H \parallel (a,b)$) and T$_\rho = 1.3$ K, respectively. The field H and temperature T of the specific heat and AC susceptibility data are divided by H$_{c2}^{C,\chi} = 0.93$ T ($\mathbf H \parallel (a,b)$) and T$_c = 0.4$ K, respectively. Solid line is the guide to the eye for the data with $\mathbf H \parallel (a,b)$. Dashed line is the guide to the eye for the data with $\mathbf H \parallel c$. Notice the same magnetic field anisotropy for the resistance- and specific heat/susceptibility-based phase diagrams.}
\label{anisotropy}
\end{figure}

In this article we present the results of transport and thermodynamic measurements of the CeIr$_{1-x}$Rh$_x$In$_5$ series for x from 0 to 0.5. The details of sample growth and characterization are described in Ref.~\onlinecite{pagliuso:prb_01}. Large  plate-like single crystals, up to 1 cm long, are grown from an excess In flux. Their quasi-2D tetragonal structure can be viewed as layers of CeIn$_3$ separated by layers of Ir$_{1-x}$Rh$_x$In$_2$. Therefore, we can treat CeIn$_3$ as the parent compound for \IrRh. Fig.~\ref{anisotropy} shows a combination of two magnetic field - temperature phase diagrams for \Ir, one based on specific heat and

susceptibility measurements that show T$_c = 0.4$K in zero field, and another based on the onset of a resistivity drop at T$_\rho = 1.3$ K in zero field. The temperature was scaled by T$_c$ and T$_\rho$ for the specific heat/susceptibility and the resistance data, respectively. The magnetic field was similarly scaled by H$_{c2}^{C,\chi} = 0.93$ T and H$_{c2}^\rho = 7$ T, critical fields for the specific heat/susceptibility and resistance for $\mathbf H \parallel (a,b)$. We can make several observation based on Fig.~\ref{anisotropy}. First, the scaled phase boundaries for $\mathbf H \parallel (a,b)$ based on resistivity and specific heat data fall on one curve (solid line) within the scatter of the data . The data for the other field orientation, $\mathbf H \parallel c$, derived from resistivity and specific heat measurements, also fall on one curve (dashed line). Therefore, the anisotropy of $\rm H_{c2}(T)$ with respect the direction of the applied magnetic field is the same for the specific heat/susceptibility and the resistance anomalies. The quasiparticles that form superconducting pairs and resistively short \Ir\ at T$_\rho$ and those that give bulk superconductivity at T$_c$ most likely have the same origin, i.e. come from the same parts of the Fermi surface. 
The absence of a resolvable anomaly in the specific heat at T$_\rho = 1.2$ K suggests that only a small part of the sample becomes superconducting at T$_\rho$. An additional hint as to the nature of the superconductivity at T$_\rho$ comes from comparing the magnitudes of the critical fields required to suppress resistive and specific heat/susceptibility anomalies to zero. The ratio of these quantities, $\rm H_{c2}^\rho / H_{c2}^{C,\chi} \approx 7.5$, suggests that the superconductivity leading to zero resistance at T$_\rho$ is filamentary in nature. It is common, for example, for the critical field of a thin film superconductor to increase when the film thickness becomes less than the superconducting coherance length (with the field in the plane of the film)~\cite{tinkham:intro-to-sc}. Following this hypothesis, let us estimate the size of the filaments. The higher critical field in filamentary superconductors is due to the reduced ability of the magnetic vortices to penetrate the filament with a cross-section smaller than the coherence length, which for \Ir\ is $\xi_{ab} \approx 240$ \AA~\cite{petrovic:epl_01}. In analogy with the case of thin film superconductivity~\cite{tinkham:intro-to-sc}, the critical field for the filamentary superconductor ${\rm H_{c2}^{film}} = {\Phi_0 \over 2\pi l \xi_{ab}}$, where $\Phi_0$ is the flux quantum and $l$ is the width of the filament. Since the critical field for bulk samples is ${\rm H_{c2}^{bulk}} = {\Phi_0 \over 2\pi \xi_{ab}^2}$, we can estimate the characteristic dimension of the filament as $l = { {{\rm H_{c2}^{bulk}} \over \rm H_{c2}^{film}} }\times \xi_{ab}  = 32$ \AA, a value typical of the size of the generic line dislocations~\cite{hirth:book,teter:pc_01}. 

The origin of such filamentary superconductivity in \Ir\ may be related to the rather high sensitivity of the bulk transition to pressure, which rises at the rate of .025 K/kbar and reaches T $ = 0.8$K at 16 kbar~\cite{sparn:nato_01}. The strain field introduced into the system by dislocations would result in creation of the superconducting regions around them at $\rm T > T_c$, shorting the sample and driving the resistance to zero. 

\begin{figure}
\epsfxsize=3in
\centerline{\epsfbox{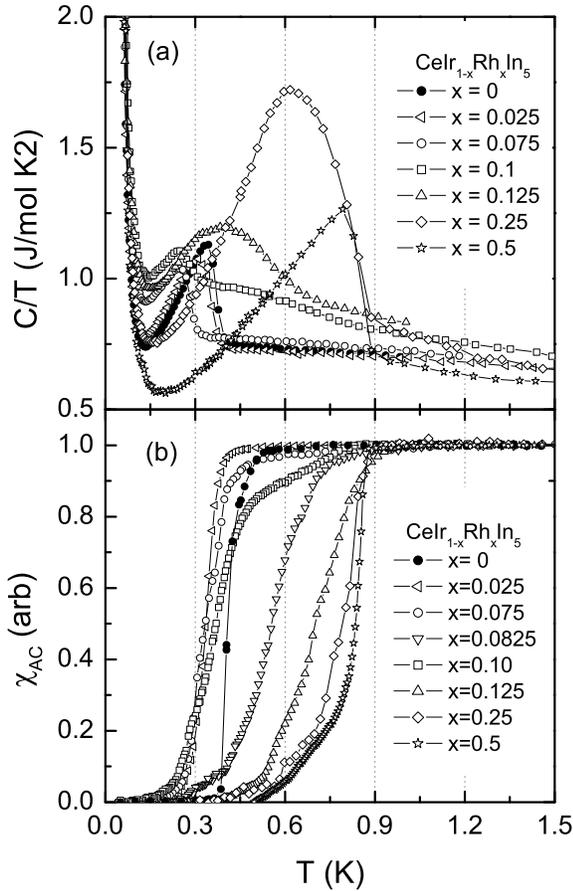}}
\caption{Specific heat (a) and AC magnetic susceptibility (b) of \IrRh\ for x from 0 to 0.5. Notice the suppression of the sharp bulk superconducting anomaly in (a) for $\rm x \leq 0.1$ and the growth of the wide anomaly at higher temperatures for $\rm \geq 0.1$. Correspondingly, the bulk of the superconducting anomaly in the susceptibility data (b) initially goes down in temperature, then broadens and moves up in temperature, and finally sharpens up again for x = 0.5.}
\label{Rh doped CeIrIn5 C and chi}
\end{figure}

The magnetic field phase diagram of resistance, specific heat, and susceptibility supports the possibility that filamentary superconductivity is the origin of the resistance anomaly at 1.2 K. This hypothesis is also consistent with the small but present sample to sample variation of the resistive anomaly, as can be seen in Fig.~\ref{R and C}: the resistance of sample \#2 does not become zero until T$ = 0.8$ K, even though the onset of the anomaly takes place at 1.3 K, similar to sample \#1 with a sharp transition to R$ = 0$ at T$ = 1.2$ K. The filamentary superconductivity around crystalline defects would indeed result in such behavior, since the density of such defects is a sample-dependent property. It is worth pointing out that within this scenario, the more perfect sample (with fewer crystalline defects) is the one with the broader resistive transition to a zero-resistance state, contrary to the usual expectation.
The filamentary superconductivity, induced by strain in the system, should be sensitive to the introduction of the crystalline defects by doping. Below we present specific heat and magnetic susceptibility data on \IrRh\ samples which bear out this expectation. Fig.~\ref{Rh doped CeIrIn5 C and chi} shows specific heat and AC magnetic susceptibility for Rh-doped samples with x between 0 and 0.5. Specific heat data, displayed in Fig.~\ref{Rh doped CeIrIn5 C and chi}(a), show two trends. First, the temperature of the sharp bulk superconducting phase transition, clearly seen for pure \Ir\ (x = 0) at T$_c = 0.4$ K, is suppressed as the Rh concentration increases from 0 to about 10\%. At the same time, a broad feature begins to rise at $\rm T > T_c$ with increasing x. The broad anomaly sharpens and moves to higher 

\noindent temperatures with x increasing from 0.125 to 0.5. In fact, the specific heat anomaly for x = 0.5 appears to be as sharp as that for x = 0. 

Similar behavior is observed in the AC susceptibility of \IrRh, shown in Fig.~\ref{Rh doped CeIrIn5 C and chi}(b). As Rh-doping is increased from zero, the anomaly moves down in temperature and broadens. At the same time the size of the drop in the susceptibility above the bulk (sharp) transition increases with x. For x above 0.1 the susceptibility feature begins to move up in temperature and sharpens again, becoming rather steep for x = 0.5, in accord with the specific heat data.

The data of Fig.~\ref{Rh doped CeIrIn5 C and chi} reflects the competition between two phenomena.  First is the suppression of the superconductivity due to the pair-breaking effect of impurities. The second one, the appearance of the broad feature in specific above the sharp bulk superconducting anomaly, and substantial broadening of the AC susceptibility anomaly, argues for inhomogeneous superconductivity in the \IrRh\ samples. Even pure \Ir\ shows the onset of small screening currents just below T$_0$ = 1.2 K, when measured with high sensitivity susceptibility apparatus~\cite{gegenwart:pc_00}. When the defect density is low, their strain fields do not overlap, and the width of the strain distribution in the sample is at the maximum: from zero in the unstrained bulk region to a maximum near the defect. Introduction of Rh impurities increases the defect density, and the volume fraction of the region of filamentary superconductivity rises with x, leading to an increase of the drop in susceptibility above the sharp bulk transition to $\approx 10$\% for the x = 0.075 sample and $\approx 80$ \% for the x = 0.1 sample. With further increase of the defect density (x $\geq$ 0.125 for \IrRh\ samples studied), the strain fields around them begin to overlap, raising the minimum strain in the sample.  The maximum strain remains unchanged, at the position of the defects. Since these are the regions of the sample where superconductivity appears first, we can expect the onset temperature of the superconductivity to be independent of x. Experimentally, the onsets of both specific heat and susceptibility anomalies for x $\geq$ 0.1 coincide at T $\approx$ 0.9 K (see Fig.~\ref{Rh doped CeIrIn5 C and chi}(b)), consistent with the above expectation. We can also understand the sharpening of both the specific heat and the susceptibility anomalies with x for x $\geq$ 0.125 as the reduction of the width of the strain distribution. Ultimately, the strain fields overlap enough to render the sample practically homogeneous with respect to strain, and the superconducting transition becomes sharp again. Apparently, we are close to this regime at x=0.5.

\begin{figure}
\epsfxsize=3in
\centerline{\epsfbox{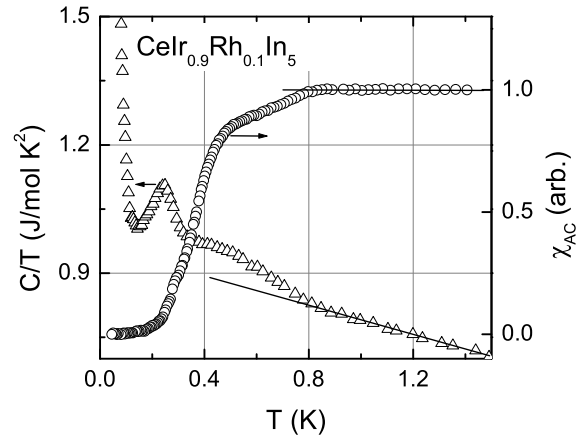}}
\caption{Specific heat C (a) and AC magnetic susceptibility $\chi_{AC}$ (b) of $ \rm CeIr_{0.9}Rh_{0.1}In_5$. Both C and $\chi_{AC}$ show broad features with onset at $\rm T \approx 0.8$ K. The sharp bulk superconducting anomaly in C at T$_c = 0.28$ K manifests itself as an additional drop in $\chi_{AC}$ at the same temperature. 80\% of the sample is screened from the  magnetic field at T$_c$.}
\label{x=0.1 sample}
\end{figure}

It is particularly instructive to look at the specific heat and susceptibility of the x = 0.1 sample, shown in Fig.~\ref{x=0.1 sample}. This sample combines the features of both clean, defect free regions of the sample, with the sharp feature at 

\noindent T$_c = 280 $ mK, and strained regions with the broad feature in specific heat above T$_c$. 
The onset of the broad feature is indicated by the kink in the specific heat data at about 0.8 K. Correspondingly, the susceptibility data also show the kink (the onset of the superconducting transition) at the same temperature. By T$_c$ the susceptibility's drop to zero is 80\% complete, and almost all of the sample is screened. There is an increase of the rate of the susceptibility drop at this temperature, reflecting the bulk nature of the specific heat feature. 

Strain-induced superconductivity in \Ir\ below T$_\rho$ is also supported by specific heat measurements of a sample that was subjected to thermal cycling while under hydrostatic pressure of 15 kbar~\cite{sparn:nato_01}. An anomaly at T $\approx$ 1 K was observed after the sample was removed from the pressure cell, whereas this anomaly was absent before pressurization. Pressurization in conjunction with thermal cycling to low temperatures apparently increased the defect density in the sample, resulting in an increase in the fraction of the sample undergoing a superconducting transition at T$_\rho$.

In conclusion, studies of \IrRh\ reveal the competition between the reduction of the superconducting T$_c$ due to the pair breaking effect and an increase in the inhomogeneous superconductivity. The detailed behavior of the specific heat and magnetic susceptibility is consistent with inhomogeneous superconductivity due to a strain field induced by crystallographic defects. The strain field becomes rather homogeneous at x = 0.5, resulting in sharp superconducting anomalies in the specific heat and susceptibility data. Our analysis of the H-T phase diagrams of \Ir\ built using the specific heat and the resistivity data suggests that the origin of the inhomogeneous superconductivity in these compounds is intrinsic filamentary superconductivity due to strain introduced by the crystallographic defects. This mechanism explains the mystery of the two superconducting anomalies (resistance and specific heat) in \Ir.

We thank I. Vekhter, R. Ramazashvili, Qimiao Si, Z. Fisk, L. Boulaevski, and D. Teter for stimulating discussions. One of us (R. M.) thanks Aspen Center for Physics for hospitality. Work at Los Alamos National Laboratory was performed under the auspices of the U.S. Department of Energy. 


\begin{thebibliography}{10}

\bibitem{petrovic:epl_01}
C. Petrovic {\it et~al.}, Europhys. Lett. {\bf 53},  354  (2001).

\bibitem{steglich79:cecu2si2}
F. Steglich {\it et~al.}, Phys. Rev. Lett. {\bf 43},  1892  (1979).

\bibitem{fisher:prl_89}
R.~A. Fisher {\it et~al.}, Phys. Rev. Lett. {\bf 62},  1411  (1989).

\bibitem{tou:jpsj_95}
H. Tou {\it et~al.}, J. Phys. Soc. Jap. {\bf 64},  725  (1995).

\bibitem{petrovic:jpcm_01}
C. Petrovic {\it et~al.}, J. Phys. Condens. Matter {\bf 13},  L337  (2001).

\bibitem{stewart:prb_01}
J.~S. Kim {\it et~al.},   , to appear in Phys. Rev. B.

\bibitem{haga:unpublished_00}
Y. Haga {\it et~al.}, unpublished  .

\bibitem{zheng:prl_01}
G. q.~Zheng {\it et~al.}, Phys. Rev. Lett. {\bf 86},  4664  (2001).

\bibitem{kohori:preprint_00}
Y. Kohori, Y. Yamato, Y. Iwamoto, and T. Kohara,   .

\bibitem{movshovich:prl_01}
R. Movshovich {\it et~al.}, Phys. Rev. Lett. {\bf 86},  5152  (2001).

\bibitem{pagliuso:prb_01}
P.~G. Pagliuso {\it et~al.}, to appear in Phys. Rev. B Rapid Commun. (2001).

\bibitem{tinkham:intro-to-sc}
M. Tinkham, {\em Introduction to Superconductivity} (McGraw Hill, New York, New
  York, 1975).

\bibitem{hirth:book}
J.~P. Hirth and J. Lothe, {\em Theory of Dislocations}, 2nd  ed. (John Wiley \&
  Sons, New York, New York, 1982).

\bibitem{teter:pc_01}
D. Teter, private communication.

\bibitem{sparn:nato_01}
G. Sparn {\it et al.}, {\it Frontiers of High Pressure Research II} (Klume), in
  press.

\bibitem{gegenwart:pc_00}
J. P. Gegenwart, private communication.

\end{thebibliography}

\end{document}